\documentclass[cits]{PoS}
\usepackage{amsmath}
\usepackage{amsthm}
\usepackage{graphicx}
\usepackage{caption}
\usepackage{subcaption}
\usepackage{multirow}
\usepackage{feynarts}
\allowdisplaybreaks
\def\Feynarts{{{\sc FeynArts}}}

\newcommand{\nn}{\pagebreak[0] \nonumber \\}
\def\aida{{{\sc Aida}}}

\usepackage{tikz}
\usetikzlibrary{"arrows", "automata", "backgrounds", "calendar", "chains", "matrix", "mindmap", "patterns", "petri", "shadows", "shapes.geometric", "shapes.misc", "spy", "trees"}
\usetikzlibrary{arrows,shapes}
\usetikzlibrary{trees}
\usetikzlibrary{matrix,arrows} 	
\usetikzlibrary{positioning}				% For "above of=" commands
\usetikzlibrary{calc,through}				% For coordinates
\usetikzlibrary{decorations.pathreplacing}
\usepackage{pgffor}

\usetikzlibrary{decorations.pathmorphing}	% For Feynman Diagrams
\usetikzlibrary{decorations.markings}
\tikzset{
    vector/.style={decorate, decoration={snake}, draw},
	provector/.style={decorate, decoration={snake,amplitude=2.5pt}, draw},
	antivector/.style={decorate, decoration={snake,amplitude=-2.5pt}, draw},
    fermion/.style={draw=black, postaction={decorate},
        decoration={markings,mark=at position .55 with {\arrow[draw=black]{>}}}},
    fermionbar/.style={draw=black, postaction={decorate},
        decoration={markings,mark=at position .55 with {\arrow[draw=black]{<}}}},
    fermionnoarrow/.style={draw=black},
    gluon/.style={decorate, draw=black,
        decoration={coil,amplitude=1.5pt, segment length=3pt}},
    scalar/.style={dashed,draw=black, postaction={decorate}},
    scalarbar/.style={dashed,draw=black, postaction={decorate},
        decoration={markings,mark=at position .55 with {\arrow[draw=black]{<}}}},
    scalarnoarrow/.style={dashed,draw=black},
    electron/.style={draw=black, postaction={decorate},
        decoration={markings,mark=at position .35 with {\arrow[draw=black]{>}}}},
    positron/.style={draw=black, postaction={decorate},
        decoration={markings,mark=at position .35 with {\arrow[draw=black]{<}}}},        
	bigvector/.style={decorate, decoration={snake,amplitude=4pt}, draw},
}
\tikzstyle{block} = [draw, rectangle, 
    minimum height=3em, minimum width=6em]
    
\newcommand{\semiloop}[4][]{%
        \draw[#1] let \p1 = ($(#3)-(#2)$) in (#3) arc (#4:({#4+180}):({0.5*veclen(\x1,\y1)});)
}

\newcommand{\valencia}{Instituto de F\'{\i}sica Corpuscular, Universitat de Val\`{e}ncia - 
Consejo Superior de Investigaciones Cient\'{\i}ficas,\\ 
Parc Cient\'{\i}fic, E-46980 Paterna, Valencia, Spain}

\title{Interplay of colour kinematics duality and analytic calculation of multi-loop scattering amplitudes: one and two loops}

\ShortTitle{Interplay of colour kinematics duality and analytic calculation of multi-loop scattering amplitudes}

\author{\speaker{William J. Torres Bobadilla}
\thanks{Talk based on a collaboration with J. Llanes, P. Mastrolia, G. Ossola, M. Passera, T. Peraro, A. Primo, G. Rodrigo and U. Schubert.}\\
\valencia\\
        E-mail: \email{william.torres@ific.uv.es}}

\abstract{
In this talk, we review recent developments towards the calculation of multi-loop scattering amplitudes.
In particular, we discuss how the colour-kinematics duality can provide new integral relations at one-loop level
via the Loop-Tree duality formalism.
On the other hand, in order to compute scattering amplitudes at one- and two-loop level,
numerically and analytically, we describe 
the preliminary automation of the adaptive integrand decomposition algorithm. 
We show preliminary results on the analytic reduction of the $\mu e$-elastic scattering at one- and two-loop level. 
}

\FullConference{13th International Symposium on Radiative Corrections (Applications of Quantum Field Theory to Phenomenology)\\
		 25-29 September, 2017 \\
		 St. Gilden, Austria}

\begin{document}

\section{Introduction}

The calculation of scattering amplitudes has been playing a very important role in the
physics of the Large Hadron Collider. Since tree-level calculations are qualitative,
higher precisions are needed in order to check the theoretical predictions with the experiments. 
Nevertheless, their calculation becomes cumbersome when increasing the higher multiplicity 
and the order in the perturbation theory. Therefore, we should take advantage of the mathematical 
properties of these scattering amplitudes.

In this talk, we discuss the new one-loop integral relations~\cite{Jurado:2017xut} that emerge as a consequence
of the Colour-Kinematics duality (CKD)~\cite{Bern:2008qj} and Loop-Tree duality formalism (LTD)~\cite{Catani:2008xa,Sborlini:2016gbr}.
Relations at integral level have been considered from a string theory perspective~\cite{Ochirov:2017jby,He:2017spx} and also from 
the modern techniques based on unitarity based methods~\cite{Chester:2016ojq,Primo:2016omk}. 

We also discuss the calculation of scattering amplitudes from the perspective of 
integrand reduction methods~\cite{Ossola:2006us,Mastrolia:2012bu,Badger:2012dv,Mastrolia:2012wf}. 
We focus our study on the automation of the recent algorithm proposed by 
Mastrolia, Peraro and Primo~\cite{Mastrolia:2016dhn}, that decomposes the space-time dimension into parallel and 
perpendicular components, $d=d_\parallel+d_\perp$. 
This decomposition allows for a straightforward classification of spurious terms
and the usual polynomial division is replaced by algebraic substitutions.
\section{Colour-Kinematics duality}
In this section, we report the results of~\cite{Jurado:2017xut}. 
For the sake of simplicity, although analogous to gauge theories coupled to matter, 
we focus on the study of the integral relations generated from
the Jacobi off-shell current $gg\to gg$. 

Let us first set up the notation and recap some features of the Colour-Kinematics duality.
The $m$-point tree-level amplitude
\begin{align}
\mathcal{A}_{m}^{\text{tree}}\left(1,2,\ldots,m\right) & =\sum_{i=1}^{N}\frac{c_{i}\,n_{i}}{D_{i}}\,\qquad D_{i}=\prod_{\alpha_{i}}s_{\alpha_{i}}\,,
\end{align}
where the sum runs over all diagrams $i$ with only cubic vertices,
$c_{i}$ are the colour factors, $n_{i}$ the kinematic numerators,
and $D_{i}$ collect the denominators of all internal propagators.

The colour factors $c_i$ obey the Jacobi identity, which in the adjoint representation, becomes
\begin{align}
-\tilde{f}^{a_{1}a_{2}x}\tilde{f}^{a_{3}a_{4}x}-\tilde{f}^{a_{1}a_{4}x}\tilde{f}^{a_{2}a_{3}x}+\tilde{f}^{a_{1}a_{3}x}\tilde{f}^{a_{2}a_{4}x} & =0\,.
\label{eq:JacExpl}
\end{align}
Therefore, we can find three colour factors, 
\begin{align}
c_{i}=\ldots\tilde{f}^{a_{1}a_{2}x}\tilde{f}^{a_{3}a_{4}x}\ldots\,, &  & c_{j}=\dots\tilde{f}^{a_{1}a_{4}x}\tilde{f}^{a_{2}a_{3}x}\ldots\,, &  & c_{k}=\ldots\tilde{f}^{a_{1}a_{3}x}\tilde{f}^{a_{2}a_{4}x}\ldots\,,
\end{align}
where the `$\ldots$' state for common terms in the three colour factors.
In such a way that the Jacobi identity is satisfied as follows,
\begin{align}
-c_{i}-c_{j}+c_{k} & =0\,.\label{eq:jac1}
\end{align}
Due to the anti-symmetry relations that colour factors and kinematic numerators satisfy
under a swapping of legs, $c_{j}\to-c_{j}\Rightarrow n_{j}\to-n_{j}$, Eq.~\eqref{eq:jac1}
can be analogously promoted to be dual in the kinematic sector,
\begin{align}
 & -n_{i}-n_{j}+n_{k}\,.\label{eq:Joff}
\end{align}
This relation between colour factors and kinematic numerators is referred
to as Colour-Kinematics duality (CKD). 

\subsection{Jacobi off-shell current}

\begin{figure}[htb!]
\centering
\includegraphics[trim = {0.5cm 0 0 4cm},clip,width=0.7\textwidth]{figs/fig2.epsi}
\caption{Jacobi combination for $gg\to gg$.}
\label{fig:JacProcesses}
\end{figure}

We generate, by following the diagrammatic approach of~\cite{Mastrolia:2015maa}, 
the off-shell current from the Jacobi
identity of the kinematic numerators of $gg\to gg$. 
This current vanishes when the
four particles attached to it are set on-shell. Nevertheless, its {\it off}-shell description
allows for a systematic study of CKD for tree higher-multiplicity or multi-loop numerators. 

We show that this off-shell currents  can be schematically represented in terms of three-point interactions.
The expressions for the Jacobi off-shell current of Fig.~\ref{fig:JacProcesses} amount to 
\begin{align}
J_{\text{g-Fey}}^{\mu_{1}\mu_{2}\mu_{3}\mu_{4}}  =\sum_{\sigma\in \mathsf{Z_{4}}}
p_{\sigma_1}^{\mu_{\sigma_1}} \Bigg(\;\parbox{20mm}{\unitlength=0.20bp%
\begin{feynartspicture}(300,300)(1,1)
\FADiagram{}
\FAProp(3.,10.)(10.,10.)(0.,){/Cycles}{0}
\FAProp(16.,15.)(10.,10.)(0.,){/Cycles}{0}
\FAProp(16.,5.)(10.,10.)(0.,){/Cycles}{0}
\FALabel(16.2273,15.5749)[cb]{\tiny $p_{\sigma_2}, \mu_{\sigma_2}$}
\FALabel(16.2273,4.5749)[ct]{\tiny $p_{\sigma_3}, \mu_{\sigma_3}$}
\FALabel(2.3,8.93)[ct]{\tiny $p_{\sigma_4}, \mu_{\sigma_4}$}
\FAVert(10.,10.){0}
\end{feynartspicture}}\;\;\Bigg)\,,
%\notag\\[-0.5cm]
\qquad
J_{\text{g-Ax}}^{\mu_{1}\mu_{2}\mu_{3}\mu_{4}}  = \sum_{\sigma\in\mathsf{ A_{4}}}
\frac{\mathcal{P}_{\sigma_{1}}^{\mu_{\sigma_{1}}\mu_{\sigma_{2}}}}{q\cdot p_{\sigma_{1}\sigma_{2}}}
q^{\alpha}\Bigg(\;\parbox{20mm}{\unitlength=0.20bp%
\begin{feynartspicture}(300,300)(1,1)
\FADiagram{}
\FAProp(3.,10.)(10.,10.)(0.,){/Cycles}{0}
\FAProp(16.,15.)(10.,10.)(0.,){/Cycles}{0}
\FAProp(16.,5.)(10.,10.)(0.,){/Cycles}{0}
\FALabel(16.2273,15.5749)[cb]{\tiny $p_{\sigma_3}, \mu_{\sigma_3}$}
\FALabel(16.2273,4.5749)[ct]{\tiny $p_{\sigma_4}, \mu_{\sigma_4}$}
\FALabel(2.3,8.93)[ct]{\tiny $-p_{\sigma_3\sigma_4}, \alpha$}
\FAVert(10.,10.){0}
\end{feynartspicture}}\;\;\Bigg)\,,
\label{eq:JgOff}
\end{align}
where $p_{ij}^\alpha\equiv\left(p_i+p_j\right)^\alpha$ and 
$\mathcal{P}_i^{\mu_{i}\mu_{j}}\equiv p_{i}^{\mu_{i}}p_{i}^{\mu_{j}}-p_{i}^{2}\,g^{\mu_{i}\mu_{j}}$.
As well, the subscripts $\text{g-Fey}$ and $\text{g-Ax}$ stand
for the Feynman and covariant part of the polarisation tensor of the gluon propagator,
\begin{align}
\Pi^{\alpha\beta}\left(p_{i},q\right)=-g^{\alpha\beta}+\frac{p_{i}^{\alpha}q^{\beta}+p_{i}^{\beta}q^{\alpha}}{p_{i}\cdot q}\,.
\label{eq:gluonprop}
\end{align}

We remark that $J_{\text{g-Fey}}$ is obtained from the Jacobi identity 
of three kinematic numerators of Eq.~\eqref{eq:Joff} 
and their structure, after algebraic manipulations, is always written
as Feynman rules where momentum conservation is not preserved. 
Likewise, CKD is straightforwardly  recovered when the four particles in the off-shell currents are set on-shell. 
On the other hand, the Feynman rules appearing in $J_{\text{g-Ax}}$ do obey momentum
conservation and the way how CKD is satisfied is individually at the level of diagrams. 

\subsection{Numerators from Jacobi off-shell currents}
Since we are interested in objects constructed from the Jacobi off-shell current~\eqref{eq:JgOff}.
We embed $J_{\text{g}}$ in a richer topology that can be either 
tree higher-multiplicity or multi-loop level,
\begin{align}
N_{\text{g}}=N_{\text{g}\,\mu_{1}\hdots\mu_{4}}X^{\mu_{1}\hdots\mu_{4}}\,,&&N_{\text{g}\,\mu_{1}\hdots\mu_{4}}=J_{\text{g}}^{\nu_{1}\hdots\nu_{4}}\Pi_{\mu_{1}\nu_{1}}\left(p_{1},q_1\right)\hdots\Pi_{\mu_{4}\nu_{4}}\left(p_{4},q_4\right)\,.
\label{eq:Numoffshell}
\end{align}
The tensors $X$ carry the information related to the kinematic part where the off-shell 
currents $J_{\text{g}}$ are embedded in.

It was noticed in~\cite{Jurado:2017xut} that in order to eliminate redundant terms in~\eqref{eq:Numoffshell},
the reference momenta $q_i$ have to be chosen to be equal for internal and external gluons. 
This request, together with the decomposition of off-shell momenta $p_i$ into massless ones, 
\begin{align}
p_{i}^{\alpha}=r_{i}^{\alpha}+\frac{p_{i}^{2}}{2q\cdot r_{i}}q^{\alpha}\,,
\label{eq:offshell}
\end{align}
rewrites the completeness relations for polarisation vectors in axial gauge as
\begin{align}
 \sum_{\lambda=1}^{d_{s}-2}\varepsilon_{\lambda\left(d_{s}\right)}^{\alpha}\left(p_{i}\right)\varepsilon_{\lambda\left(d_{s}\right)}^{*\beta}\left(p_{i}\right)
 &=\sum_{\lambda_{i}=1}^{d_{s}-2}\varepsilon_{i}^{\alpha}\varepsilon_{i}^{*\beta}+\frac{p_{i}^{2}}{\left(r_{i}\cdot q\right)^{2}}q^{\alpha}q^{\beta}
=-g^{\alpha\beta}+\frac{r_{i}^{\alpha}q^{\beta}+r_{i}^{\beta}q^{\alpha}}{r_{i}\cdot q}+\frac{p_{i}^{2}}{\left(r_{i}\cdot q\right)^{2}}q^{\alpha}q^{\beta}\,,
 \label{eq:gluonCR}
 \end{align}
allowing to distinguish between
on- and off-shell quantities. The latter takes care of contributions coming from $p_i^2$ only. 

Therefore, Eq.~\eqref{eq:Numoffshell} becomes, 
\begin{align}
N_{\text{g}}^{\nu_{1}\hdots\nu_{4}}& = \frac{1}{2}\sum_{i,j,k,l=1}^{4}\epsilon_{ijkl}\,p_{i}^{2}\left(A_{ijkl}\,\mathcal{E}_{ij}^{\nu_{i}\nu_{j}}\mathcal{E}_{kl}^{\nu_{k}\nu_{l}}
+B_{ijkl}\,\mathcal{E}_{jk}^{\nu_{j}\nu_{k}}\mathcal{Q}_{l}^{\nu_{i}\nu_{l}}
+C_{ijkl}\,p_{j}^{2}\,\mathfrak{q}^{\nu_{i}\nu_{j}}\mathcal{E}_{kl}^{\nu_{k}\nu_{l}}\right)\,,
\label{eq:JgOffDec}
\end{align}
where the functions $A, B$ and $C$, because of the way the off-shell wave-functions are parametrised, 
do not contain a dependence on $p_i^2$. 
Also, in the parametric form of $N_{\text{g}}$ no terms proportional to $p_i^2p_j^2p_k^2p_l^2$ or $p_i^2p_j^2p_k^2$
are present. 
This is indeed due to the choice of a unique reference momentum $q$ in the definition of 
internal propagators and wave-functions. Hence, any numerator built from the off-shell current 
$J_{\text{g}}$ is written in terms of (at most)
the product of two squared momenta, $p_i^2p_j^2$. 

\subsection{One-loop integral relations}
We provide one-loop integral relations that are obtained by combining
CKD with the Loop-Tree duality (LTD) formalism. 
It turns out that with the numerator 
\begin{align}
\includegraphics[trim = {2.5cm 2.5cm 0 0},clip,width=0.8\textwidth]{figs/fig3.epsi}
\end{align}
we have, from the decomposition~\eqref{eq:JgOffDec}, 
\begin{align}
I_3 &= \int_{\ell} \frac{N_3}{\ell^2(\ell+p_3)^2(\ell+p_{34})^2}\nn
&=\int_{\ell}\left\{
\tilde{A}_{11}\,\mathcal{I}\left[\!\!\!\parbox{25mm}{\begin{tikzpicture}[line width=1 pt,node distance=0.5 cm and 0.5 cm]
\coordinate[] (v1);
\coordinate[left = of v1, label= left :\footnotesize$p_1$] (p1);
\coordinate[right = of v1] (v2);
\coordinate[above right = of v2, label= right :\footnotesize$p_2$] (p2);
\coordinate[right = of v2, label= right :\footnotesize$p_3$] (p3);
\coordinate[below right = of v2, label= right :\footnotesize$p_4$] (p4);

\draw[fermionnoarrow] (p1) -- (v1);
\semiloop[fermion]{v1}{v2}{0};
\semiloop[fermionnoarrow]{v2}{v1}{180};
\draw[fermionnoarrow] (p2) -- (v2);
\draw[fermionnoarrow] (p3) -- (v2);
\draw[fermionnoarrow] (p4) -- (v2);

%\draw[fill=black] (cent) circle (.05cm);
\end{tikzpicture}}\,\right] 
+ \tilde{A}_{12}\,\mathcal{I}\left[\!\!\!\parbox{25mm}{\begin{tikzpicture}[line width=1 pt,node distance=0.5 cm and 0.5 cm]
\coordinate[] (v1);
\coordinate[left = of v1, label= left :\footnotesize$p_2$] (p1);
\coordinate[right = of v1] (v2);
\coordinate[above right = of v2, label= right :\footnotesize$p_3$] (p2);
\coordinate[right = of v2, label= right :\footnotesize$p_4$] (p3);
\coordinate[below right = of v2, label= right :\footnotesize$p_1$] (p4);

\draw[fermionnoarrow] (p1) -- (v1);
\semiloop[fermionnoarrow]{v1}{v2}{0};
\semiloop[fermion]{v2}{v1}{180};
\draw[fermionnoarrow] (p2) -- (v2);
\draw[fermionnoarrow] (p3) -- (v2);
\draw[fermionnoarrow] (p4) -- (v2);

%\draw[fill=black] (cent) circle (.05cm);
\end{tikzpicture}}\,\right]  
+ \tilde{C}_{11}\,\mathcal{I}\left[\parbox{15mm}{\begin{tikzpicture}[line width=1 pt,node distance=0.25 cm and 0.5 cm]
\coordinate[] (v1);
\coordinate[right = of v1] (v2);
\coordinate[right = of v2] (v3);

\coordinate[above = of v3, label= right :\footnotesize$p_3$] (p3);
\coordinate[above = of p3, label= right :\footnotesize$p_2$] (p2);
\coordinate[below = of v3, label= right :\footnotesize$p_4$] (p4);
\coordinate[below = of p4, label= right :\footnotesize$p_1$] (p1);

\semiloop[fermionnoarrow]{v1}{v2}{0};
\semiloop[fermion]{v2}{v1}{180};
\draw[fermionnoarrow] (p1) -- (v2);
\draw[fermionnoarrow] (p2) -- (v2);
\draw[fermionnoarrow] (p3) -- (v2);
\draw[fermionnoarrow] (p4) -- (v2);

%\draw[fill=black] (cent) circle (.05cm);
\end{tikzpicture}}\,\right]\right\} \,,
\end{align}
where $\mathcal{I}[\cdots]$ states for the integrand of the topology, $\tilde{A}$ and $\tilde{C}$ are polynomials
in the loop momentum $\ell$.

In the case of a massless theory, 
we find relations between integrals with the same number 
of propagators. For the $2\to2$ case, we  write relations for Feynman integrals 
with three loop propagators. In~\cite{Jurado:2017xut}, we study 
particular example
of $gg\to ss$ showing that higher rank numerators can be replaced by lower ones. 
This outcome indeed allows for an optimisation in the evaluation of  Feynman integrals.

\section{Adaptive Integrand Decomposition}

In this section, we explain the main features of the 
Adaptive Integrand Decomposition Algorithm (\aida), the automation
of the recent method proposed by Mastrolia, Primo and Peraro~\cite{Mastrolia:2016dhn}.
We remark that this method decomposes the space-time dimension, $d=4-2\epsilon$, into parallel
(or longitudinal) and orthogonal (or transverse) dimensions,
$d=d_{\parallel}+d_{\perp}$.
Parallel and orthogonal directions show particular properties for
topologies with less than five external legs. 

In the structure of the Feynman integrals,
\begin{align}
\mathcal{I}_{i_{1}\cdots i_{n}}^{(\ell)}[\mathcal{N}] & =\int\left(\prod_{i=1}^{\ell}\frac{d^{d}\bar{l}_{i}}{\pi^{d/2}}\right)\frac{\mathcal{N}_{i_{1}\cdots i_{n}}(\bar{l}_{i})}{\prod_{j}D_{j}(\bar{l}_{j})}\,,
\label{eq:hloopsInt}
\end{align}
loop momenta become
\begin{align}
\bar{l}_{i}^{\alpha}=l_{\parallel\,i}^{\alpha}+\lambda_{i}^{\alpha}\,,\label{eq:newdeco}
\end{align}
with 
\begin{align}
 & \bar{l}_{\parallel\,i}^{\alpha}=\sum_{j=1}^{d_{\parallel}}x_{ji}\,e_{j}^{\alpha}\,, &  & \lambda_{i}^{\alpha}=\sum_{j=d_{\parallel}+1}^{4}x_{ji}\,e_{j}^{\alpha}+\mu_{i}^{\alpha}\,, &  & \lambda_{ij}=\sum_{l=d_{\parallel}+1}^{4}x_{li}\,x_{lj}+\mu_{ij}\,.
\end{align}
In Eq.~\eqref{eq:newdeco}, $l_{\parallel\,i}$ is a vector of the $d_{\parallel}$-dimensional
space spanned by the external momenta, and $\lambda_{i}$ belongs
the $d_{\perp}$-dimensional orthogonal subspace. 
In this parametrisation, all denominators become independent of the transverse components of
the loop momenta.

Let us indicate with $\mathbf{z}$ the full set of $\ell(\ell+9)/2$
variables 
\begin{align}
\mathbf{z}= & \{\mathbf{x}_{\parallel\,i},\mathbf{x}_{\perp\,i},\lambda_{ij}\},\quad i,j=1,\dots\ell\,,
\end{align}
where $\mathbf{x}_{\parallel\,i}$ ($\mathbf{x}_{\perp\,i}$) are
the components of the loop momenta parallel (orthogonal) to the external
kinematics, the denominators are reduced to polynomials in the subset
of variables 
\begin{align}
\boldsymbol{\tau}= & \{\mathbf{x}_{\parallel},\lambda_{ij}\},\quad\boldsymbol{\tau}\subset\mathbf{z},
\end{align}
so that the general $r$-point integrand has the form 
\begin{align}
\mathcal{I}_{i_{1}\dots i_{r}}(\boldsymbol{\tau},\mathbf{x}_{\perp})\equiv\frac{\mathcal{N}_{i_{1}\dots i_{r}}(\boldsymbol{\tau},\mathbf{x}_{\perp})}{D_{i_{1}}(\boldsymbol{\tau})\cdots D_{i_{r}}(\boldsymbol{\tau})}\,.
\end{align}
Since numerator and denominators depend on different variables, 
the adaptive integrand decomposition suggests the following algorithm:
\begin{enumerate}
\item \textbf{Divide:} we divide the numerator $\mathcal{N}_{i_{1}\dots i_{r}}(\boldsymbol{\tau},\mathbf{x}_{\perp})$
modulo the Gr\"obner basis $\mathcal{G}_{i_{1}\cdots i_{r}}(\boldsymbol{\tau})$
of the ideal $\mathcal{J}_{i_{1}\cdots i_{r}}(\boldsymbol{\tau})$
generated by the set of denominators. The polynomial division is performed
be adopting the lexicographic ordering $\lambda_{ij}\ll\mathbf{x}_{\parallel}$,
\begin{align}
\mathcal{N}_{i_{1}\dots i_{r}}(\boldsymbol{\tau},\mathbf{x}_{\perp})=\sum_{k=1}^{r}\mathcal{N}_{i_{1}\dots i_{k-1}i_{k+1}\dots i_{r}}(\boldsymbol{\tau},\mathbf{x}_{\perp})D_{i_{k}}(\boldsymbol{\tau})+\Delta_{i_{1}\dots i_{r}}(\mathbf{x}_{\parallel},\mathbf{x}_{\perp})\,.
\end{align}
The Gr\"obner basis does not need to be explicitly computed, since, with the choice
of variables and the ordering described here, the division is equivalent
to applying the set of linear relations described above.

\item \textbf{Integrate:} Since denominators do not depend on transverse variables,
$\mathbf{x}_{\perp}$, we can integrate the residue $\Delta_{i_{1}\dots i_{r}}$
over transverse directions. This integration is carried out by expressing
$\Delta_{i_{1}\dots i_{r}}$ in terms of Gegenbauer polynomials, i.e.,
\begin{align}
\Delta_{i_{1}\dots i_{r}}^{\text{int}}(\boldsymbol{\tau})=\int\!d^{(4-d_{\parallel})\ell}\boldsymbol{\Theta}_{\perp}\Delta_{i_{1}\dots i_{r}}(\boldsymbol{\tau},\boldsymbol{\Theta}_{\perp})\,.
\end{align}
Where $\Delta_{i_{1}\dots i_{r}}^{\text{int}}$ is a polynomial in
$\boldsymbol{\tau}$ whose coefficients depend on the space-time dimension
$d$. 
\item \textbf{Divide:} the structure of the integrated residue suggests a second
division. This can be seen from the dependence $\Delta_{i_{1}\dots i_{r}}^{\text{int}}$
has on the variables $\boldsymbol{\tau}$. In fact, after applying
the division, similarly as in the first step of this algorithm, we
get 
\begin{align}
\Delta_{i_{1}\dots i_{r}}^{\text{int}}(\boldsymbol{\tau})=\sum_{k=1}^{r}\mathcal{N}_{i_{1}\dots i_{k-1}i_{k+1}\dots i_{r}}^{\text{int}}(\boldsymbol{\tau})D_{i_{k}}(\boldsymbol{\tau})+\Delta_{i_{1}\dots i_{r}}^{\prime}(\mathbf{x}_{\parallel}),
\end{align}
where the new residue $\Delta_{i_{1}\dots i_{r}}^{\prime}(\mathbf{x}_{\parallel})$
can only depend on $\mathbf{x}_{\parallel}$.
\end{enumerate}
Apart from the examples provided in~\cite{Mastrolia:2016dhn}, 
this algorithm has also been applied to the leading color contribution to the two-loop all-plus five- gluon amplitude~\cite{Mastrolia:2016czu,TorresBobadilla:2017kpd}.

\subsection{$\mu e$ elastic scattering}
Motivated by the new experiment MUonE proposed at CERN~\cite{Abbiendi:2016xup} that provides 
a new and independent determination of the leading hadronic contribution to the muon 
$g$-2~\cite{Abbiendi:2016xup,Calame:2015fva}, 
we consider as an application of \aida,
the one- and two-loop reductions of the $\mu e$ elastic scattering,
\begin{align}
e^{-}\left(-p_{1}\right)\,\mu^{-}\left(-p_{4}\right) & \to e^{-}\left(p_{2}\right)\,\mu^{-}\left(p_{3}\right)\,,
\end{align}
The electron is treated as massless, $m_{e}^{2}=0$, while we retain full dependence on the muon mass, $m_{\mu}^{2}\ne0$. 
We focus on the spin summed/averaged squared matrix elements,
\begin{align}
M^{\left(1\right)}=2\text{Re}\langle A_{e\mu}^{\left(0\right)}|A_{e\mu}^{\left(1\right)}\rangle\,, &  &  & M^{\left(2\right)}=2\text{Re}\langle A_{e\mu}^{\left(0\right)}|A_{e\mu}^{\left(2\right)}\rangle\,,
\end{align}
to obtain the results of the corresponding one- and two-loop amplitudes. 

Hence, we define the kinematical variables this amplitude depends
on to be
\begin{align}
s=\left(p_{1}+p_{2}\right)^{2}\,, &  & t=\left(p_{2}+p_{3}\right)^{2}\,, &  & u=\left(p_{1}+p_{3}\right)^{2}=-s-t+2m_{\mu}^{2}\,.
\end{align}

\subsubsection{One-loop}

\begin{figure}[htb!]
\centering
\includegraphics[scale=0.6]{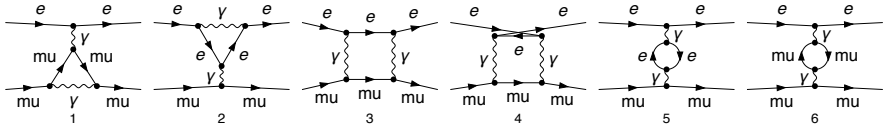}
\caption{Feynman diagrams contributing to the one-loop $\mu e$-scattering amplitude.}
\label{fig:emu1L}
\end{figure}

In order to compute the one-loop amplitude, we have to evaluate the
Feynman diagrams of Fig.~\ref{fig:emu1L}. The algorithm we
use to perform the integrand reduction by means of \aida~is described
as follows\footnote{Further details will be provided in: P. Mastrolia, T. Peraro, A. Primo and W. J. Torres Bobadilla (in preparation).}
\begin{enumerate}
\item Identify parent topologies and group diagrams: according to the numbering
of diagrams of Fig.~\ref{fig:emu1L}, we end up with three groups,
\begin{align}
\{\{3,2,5\},\{4\},\{1,6\}\}\,,
\end{align}
 being the first element of each sublist the parent topology.
\item Generate cuts: the parent topology of the group $1$
\begin{align}
&\parbox{20mm}{\includegraphics[scale=0.2]{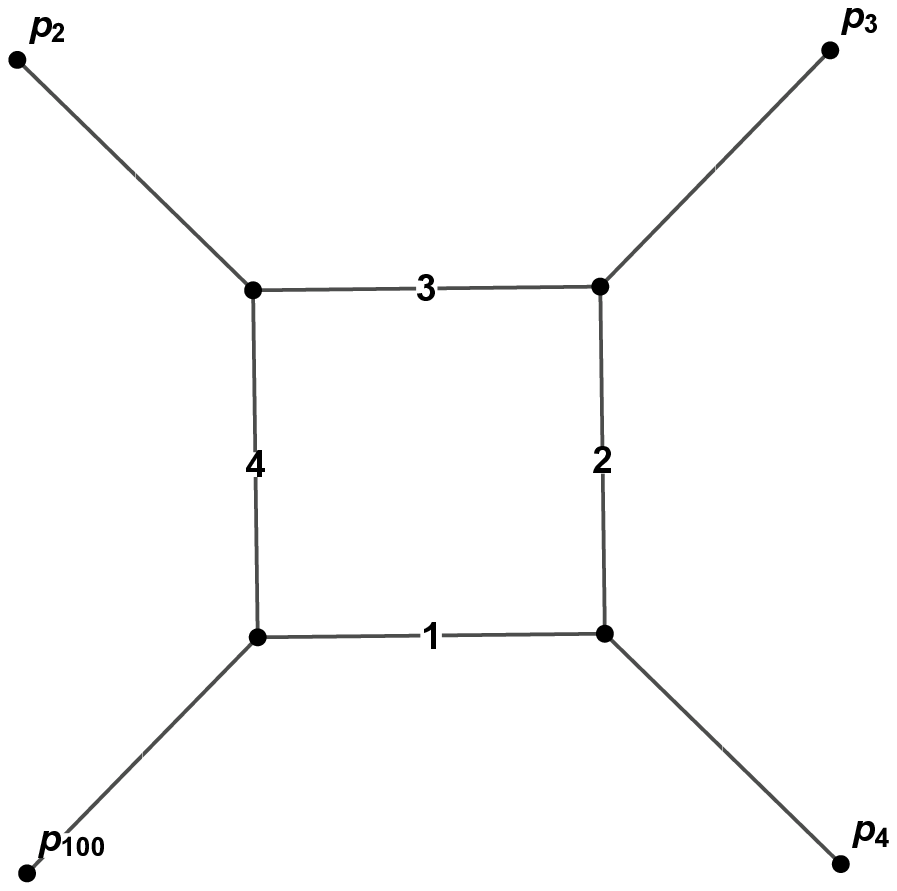}}
\end{align}
generates
\begin{align}
 & \{\left(1234\right),\left(234\right),\left(134\right),\left(124\right),\left(123\right),\left(34\right),
 %\nonumber \\
 %&
  \left(24\right),\left(23\right),\left(14\right),\left(13\right),\left(12\right),\left(4\right),\left(3\right),\left(2\right),\left(1\right)\}\,.
\end{align}
\item Define adaptive variables and prepare substitution rules for all cuts:
for instance, the parametrisation of the diagram $5$ of Fig.~\ref{fig:emu1L}
becomes\footnote{We have, for simplicity, set $s=1$. 
Nevertheless, its dependence can be recovered from dimensional analysis.}
\begin{align}
x_{1}^{\left(13\right)} & \to\frac{d_{1}}{2}-\frac{d_{3}}{2}+\frac{1}{2}\,,&
\lambda_{11}^{\left(13\right)} & \to-\frac{d_{1}^{2}}{4}+\left(\frac{d_{3}}{2}+\frac{1}{2}\right)d_{1}+\frac{d_{3}}{2}-\frac{d_{3}^{2}}{4}-\frac{1}{4}\,.
\end{align}
\item Organise cuts in jobs: let us illustrate this organisation by considering
the tadpole contribution to the amplitude coming from the first group.
For instance, assigning the number $\left(2\right)$ to the massive
propagator of the parent topology, we end up with
\begin{align}
 & \left\{ \text{N}_{\left(2\right)}^{\left(1234\right)},\text{N}_{\left(2\right)}^{\left(234\right)},\text{N}_{\left(2\right)}^{\left(24\right)}\right\} \,,
\end{align}
where the first term corresponds to the tadpole contribution coming
from the reduction of the Feynman diagram $3$ and the other contributions
represent triangle and bubble reductions, diagrams $2$ and $5$ respectively.
\item Divide: apply substitution rules of 3. to the numerator.
\item Collect powers of denominators to read off residue and numerators
of lower cuts.
\item Integrate (substitute) transverse variables appearing in the residues.
\item Divide again, using as input numerators the residues.
\end{enumerate}
After performing these steps, we obtain a decomposition in terms of
one-loop scalar integrals. For sake of simplicity we do not write
the complete expression, moreover, it is available in an ancillary
included in the \verb"arXiv" submission, containing the coefficients
which follows the notation of the FeynCalc package~\cite{Mertig:1990an,Shtabovenko:2016sxi}.

\subsubsection{Two-loop}
\begin{figure}[htb!]
\centering
\includegraphics[scale=0.55]{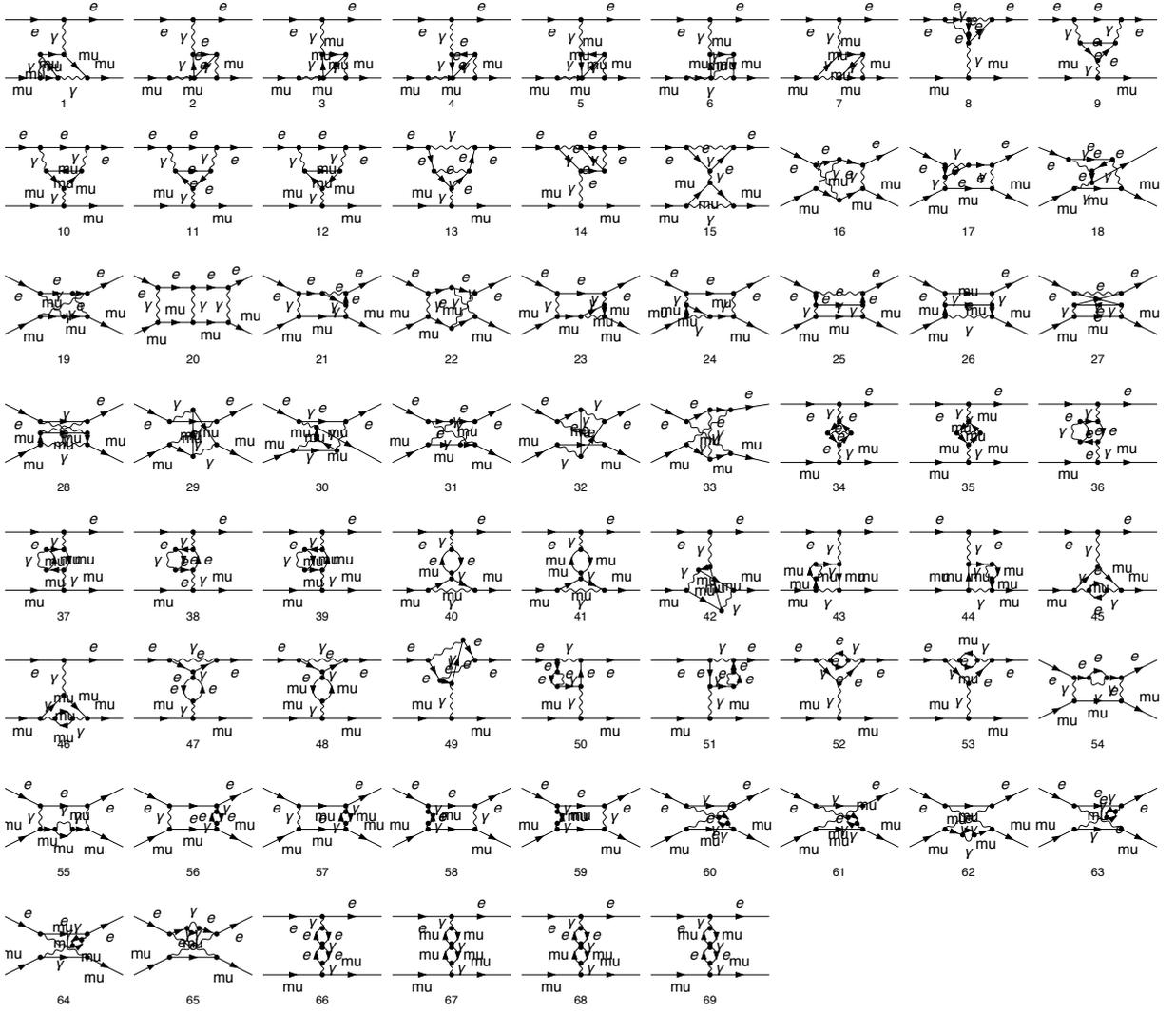}
\caption{Feynman diagrams contributing to the two-loop $\mu e$-scattering amplitude.}
\label{fig:emu2L}
\end{figure}

For the calculation of the two-loop amplitude of the $\mu e$ elastic scattering, we have to evaluate 
the Feynman diagrams of Fig.~\ref{fig:emu2L}. 
We follow the very same procedure as the one described in the previous section.  
Nevertheless, we describe the steps that could give a different interpretation
\begin{enumerate}
\item Identify parent topologies and group diagrams:
\begin{align}
\big\{&\{1,46\},\{3,5\},\{6,35,41,67\},\{7\},\{16\},\{17,8,36,38,50,52,54,58\},\{18,60,65\},\{19\},\nn
&\{20\},\{21,14,51,56\},\{22\},\{23,57\},\{24,55,59\},\{25,2,4,9,11,13,34,47,66\},\nn
&\{26,10,12,15,37,39,40,43,44,48,68,69\},\{27\},\{28\},\{29,61,62\},\{30,64\},\{31\},\nn
&\{32,63\},\{33\},\{42\},\{45\},\{49\},\{53\}\big\}\,.
\end{align}
\item Generate cuts: let us focus on the sixth group, whose parent topology
\begin{align}
&\parbox{20mm}{\includegraphics[scale=0.35]{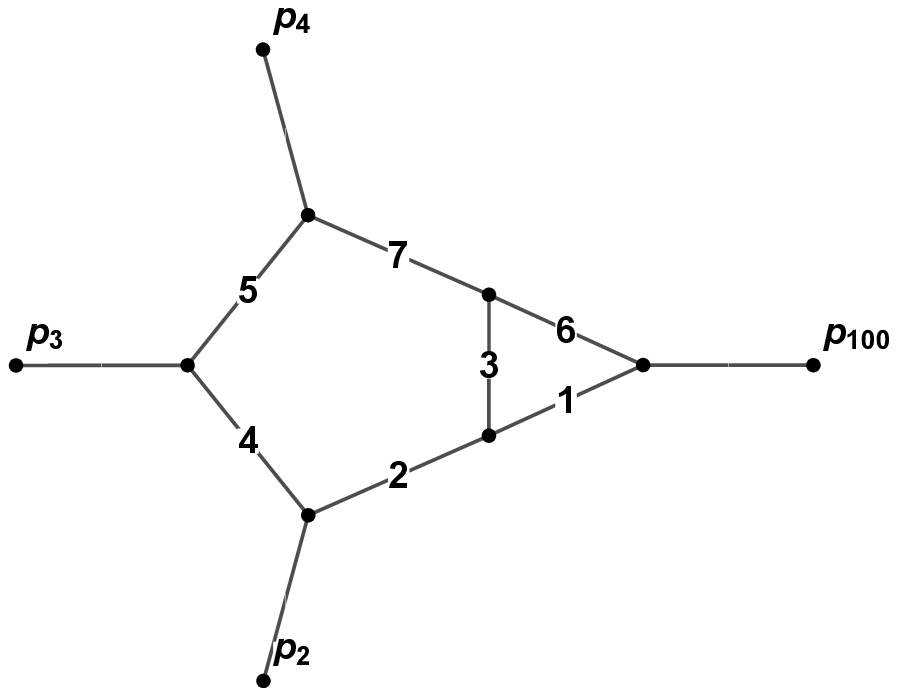}} 
\end{align}
gives contribution to the following cuts
\begin{align}
\big\{&\{\{1,6\},\{2,4,5,7\},\{3\}\},\{\{6\},\{2,4,5,7\},\{3\}\},\{\{1,6\},\{4,5,7\},\{3\}\},\{\{1,6\},\{2,4,5,7\}\},\nn
&\{\{1,6\},\{2,5,7\},\{3\}\},\{\{1,6\},\{2,4,7\},\{3\}\},\{\{1\},\{2,4,5,7\},\{3\}\},\{\{1,6\},\{2,4,5\},\{3\}\},\nn
&\{\{6\},\{4,5,7\},\{3\}\},\{\{6\},\{2,4,5,7\}\},\{\{6\},\{2,5,7\},\{3\}\},\{\{6\},\{2,4,7\},\{3\}\},\{\{2,4,5,7\},\nn
&\{3\}\},\{\{6\},\{2,4,5\},\{3\}\},\{\{1,6\},\{4,5,7\}\},\hdots,\{\{1\},\{3\}\},\{\{1\},\{2\}\}\big\}\,.
\end{align}
  \setcounter{enumi}{3}
  \item Organise cuts in jobs: in order to consider the full contribution of all possible diagrams, 
  there is a subtlety w.r.t. the one-loop case, which relies on the treatment of the diagrams
  that contain squared propagators, e.g. diagrams 51, 52, 53 in~Fig.~\ref{fig:emu2L}.
  For instance, the contribution to the cut $\{\{1,6\},\{2,4,5\},\{3\}\}$ with linear propagators  gets contributions
  from
\begin{align}
\left\{ \text{N}_{\{\{1,6\},\{2,5,7\},\{3\}\}}^{\{\{1,6\},\{2,4,5,7\},\{3\}\}},\text{N}_{\{\{1,6\},\{2,5,7\},\{3\}\}}^{\{\{1,6\},\{\dot{2},5,7\},\{3\}\}},\text{N}_{\{\{1,6\},\{2,5,7\},\{3\}\}}^{\{\{1,6\},\{2,5,\dot{7}\},\{3\}\}}\right\}  \,,
\end{align}
where dots represent squared propagators.
\end{enumerate}
Unless the one-loop case, the two-loop result \aida~provides yet needs the evaluation
of the Feynman integrals or reduce them to master integrals by means of 
Integration-by-parts (IBPs) identities~\cite{Tkachov:1981wb,Chetyrkin:1981qh,Laporta:2001dd}.
Nevertheless, the calculation of these master integrals has been carried out for the 
planar families~\cite{Mastrolia:2017pfy} (see references therein). 

In order to illustrate \aida's output, we include in the \verb"arXiv" submission, 
the list of irreducible polynomials.

\section{Conclusions}
We have studied the colour-kinematics duality (CKD) at tree and one-loop level.
For the former, we have studied the features of the Jacobi off-shell currents 
when they generate tree higher-multiplicity or multi-loop objects. 
Whereas, for the latter, together with the Loop-Tree duality formalism, integral relations among 
numerators constructed from CKD have been considered. 

For the evaluation of one- and two-loop scattering amplitudes, 
we have described the preliminary implementation of the 
Adaptive Integrand Decomposition Algorithm (\aida). 
This implementation automates, numerically and analytically, the method of~\cite{Mastrolia:2016dhn}.
We  have also shown results for the analytic reduction of the one- and two-loop amplitudes of
the $\mu e$-elastic scattering.

\acknowledgments
This work is supported by the Spanish
Government and ERDF funds from European Commission (Grants No. FPA2014-53631-C2-1-P and SEV-2014-
0398) and by Consejo Superior de Investigaciones Cient\'{\i}ficas (Grant No. PIE-201750E021). 

The Feynman diagrams depicted in this paper were generated using \Feynarts~\cite{Hahn:2000kx}.

\bibliographystyle{JHEP-LL12}
\bibliography{refs}

\providecommand{\href}[2]{#2}\begingroup\raggedright\begin{thebibliography}{10}

\bibitem{Jurado:2017xut}
J.~L. Jurado, G.~Rodrigo, and W.~J. Torres~Bobadilla {\em JHEP} {\bf 12} (2017)
  122, [\href{http://xxx.lanl.gov/abs/1710.11010}{{\tt 1710.11010}}].

\bibitem{Bern:2008qj}
Z.~Bern, J.~Carrasco, and H.~Johansson {\em Phys.Rev.} {\bf D78} (2008) 085011,
  [\href{http://xxx.lanl.gov/abs/0805.3993}{{\tt 0805.3993}}].

\bibitem{Catani:2008xa}
S.~Catani, T.~Gleisberg, F.~Krauss, G.~Rodrigo, and J.-C. Winter {\em JHEP}
  {\bf 09} (2008) 065, [\href{http://xxx.lanl.gov/abs/0804.3170}{{\tt
  0804.3170}}].

\bibitem{Sborlini:2016gbr}
G.~F.~R. Sborlini, F.~Driencourt-Mangin, R.~Hernandez-Pinto, and G.~Rodrigo
  {\em JHEP} {\bf 08} (2016) 160,
  [\href{http://xxx.lanl.gov/abs/1604.06699}{{\tt 1604.06699}}].

\bibitem{Ochirov:2017jby}
A.~Ochirov, P.~Tourkine, and P.~Vanhove
  \href{http://xxx.lanl.gov/abs/1707.05775}{{\tt 1707.05775}}.

\bibitem{He:2017spx}
S.~He, O.~Schlotterer, and Y.~Zhang
  \href{http://xxx.lanl.gov/abs/1706.00640}{{\tt 1706.00640}}.

\bibitem{Chester:2016ojq}
D.~Chester {\em Phys. Rev.} {\bf D93} (2016), no.~6 065047,
  [\href{http://xxx.lanl.gov/abs/1601.00235}{{\tt 1601.00235}}].

\bibitem{Primo:2016omk}
A.~Primo and W.~J. Torres~Bobadilla {\em JHEP} {\bf 04} (2016) 125,
  [\href{http://xxx.lanl.gov/abs/1602.03161}{{\tt 1602.03161}}].

\bibitem{Ossola:2006us}
G.~Ossola, C.~G. Papadopoulos, and R.~Pittau {\em Nucl. Phys.} {\bf B763}
  (2007) 147--169, [\href{http://xxx.lanl.gov/abs/hep-ph/0609007}{{\tt
  hep-ph/0609007}}].

\bibitem{Mastrolia:2012bu}
P.~Mastrolia, E.~Mirabella, and T.~Peraro {\em JHEP} {\bf 06} (2012) 095,
  [\href{http://xxx.lanl.gov/abs/1203.0291}{{\tt 1203.0291}}]. [Erratum:
  JHEP11,128(2012)].

\bibitem{Badger:2012dv}
S.~Badger, H.~Frellesvig, and Y.~Zhang {\em JHEP} {\bf 1208} (2012) 065,
  [\href{http://xxx.lanl.gov/abs/1207.2976}{{\tt 1207.2976}}].

\bibitem{Mastrolia:2012wf}
P.~Mastrolia, E.~Mirabella, G.~Ossola, and T.~Peraro {\em Phys. Rev.} {\bf D87}
  (2013), no.~8 085026, [\href{http://xxx.lanl.gov/abs/1209.4319}{{\tt
  1209.4319}}].

\bibitem{Mastrolia:2016dhn}
P.~Mastrolia, T.~Peraro, and A.~Primo {\em JHEP} {\bf 08} (2016) 164,
  [\href{http://xxx.lanl.gov/abs/1605.03157}{{\tt 1605.03157}}].

\bibitem{Mastrolia:2015maa}
P.~Mastrolia, A.~Primo, U.~Schubert, and W.~J. Torres~Bobadilla {\em Phys.
  Lett.} {\bf B753} (2016) 242--262,
  [\href{http://xxx.lanl.gov/abs/1507.07532}{{\tt 1507.07532}}].

\bibitem{Mastrolia:2016czu}
P.~Mastrolia, T.~Peraro, A.~Primo, and W.~J. Torres~Bobadilla {\em PoS} {\bf
  LL2016} (2016) 007, [\href{http://xxx.lanl.gov/abs/1607.05156}{{\tt
  1607.05156}}].

\bibitem{TorresBobadilla:2017kpd}
W.~J. Torres~Bobadilla, {\em {Generalised Unitarity, Integrand Decomposition,
  and Hidden properties of QCD Scattering Amplitudes in Dimensional
  Regularisation}}.
\newblock PhD thesis, Padua U., 2017.

\bibitem{Abbiendi:2016xup}
G.~Abbiendi {\em et~al.} {\em Eur. Phys. J.} {\bf C77} (2017), no.~3 139,
  [\href{http://xxx.lanl.gov/abs/1609.08987}{{\tt 1609.08987}}].

\bibitem{Calame:2015fva}
C.~M. Carloni~Calame, M.~Passera, L.~Trentadue, and G.~Venanzoni {\em Phys.
  Lett.} {\bf B746} (2015) 325--329,
  [\href{http://xxx.lanl.gov/abs/1504.02228}{{\tt 1504.02228}}].

\bibitem{Mertig:1990an}
R.~Mertig, M.~Bohm, and A.~Denner {\em Comput. Phys. Commun.} {\bf 64} (1991)
  345--359.

\bibitem{Shtabovenko:2016sxi}
V.~Shtabovenko, R.~Mertig, and F.~Orellana
  \href{http://xxx.lanl.gov/abs/1601.01167}{{\tt 1601.01167}}.

\bibitem{Tkachov:1981wb}
F.~V. Tkachov {\em Phys. Lett.} {\bf B100} (1981) 65--68.

\bibitem{Chetyrkin:1981qh}
K.~G. Chetyrkin and F.~V. Tkachov {\em Nucl. Phys.} {\bf B192} (1981) 159--204.

\bibitem{Laporta:2001dd}
S.~Laporta {\em Int. J. Mod. Phys.} {\bf A15} (2000) 5087--5159,
  [\href{http://xxx.lanl.gov/abs/hep-ph/0102033}{{\tt hep-ph/0102033}}].

\bibitem{Mastrolia:2017pfy}
P.~Mastrolia, M.~Passera, A.~Primo, and U.~Schubert {\em JHEP} {\bf 11} (2017)
  198, [\href{http://xxx.lanl.gov/abs/1709.07435}{{\tt 1709.07435}}].

\bibitem{Hahn:2000kx}
T.~Hahn {\em Comput. Phys. Commun.} {\bf 140} (2001) 418--431,
  [\href{http://xxx.lanl.gov/abs/hep-ph/0012260}{{\tt hep-ph/0012260}}].

\end{thebibliography}\endgroup

\end{document}